\documentclass[prb,twocolumn,superscriptaddress]{revtex4}

\usepackage{graphicx}
\usepackage{amssymb,amsmath}
\usepackage{dcolumn}
\usepackage{bm}
\usepackage{color}

\begin{document}

\title{One-dimensional Van Hove polaritons}

\author{K. B. Arnardottir}
\affiliation{Science Institute, University of Iceland, Dunhagi-3,
IS-107, Reykjavik, Iceland}
\affiliation{Fysikum, Stockholms Universitet, S-106 91 Stockholm, Sweden}

\author{O. Kyriienko}
\affiliation{Science Institute, University of Iceland, Dunhagi-3,
IS-107, Reykjavik, Iceland}
\affiliation{Division of Physics and Applied Physics, Nanyang Technological University 637371, Singapore}

\author{M. E. Portnoi}
\affiliation{School of Physics, University of Exeter, Stocker Road, Exeter EX4 4QL, United Kingdom}
\affiliation{International Institute of Physics, Avenida Odilon Gomes de Lima, 1722, Capim Macio, C\'odigo de Endere\c{c}amento Postal: 59078-400, Natal - RN, Brazil}

\author{I. A. Shelykh}
\affiliation{Science Institute, University of Iceland, Dunhagi-3,
IS-107, Reykjavik, Iceland} \affiliation{Division of Physics and
Applied Physics, Nanyang Technological University 637371, Singapore}

\date{\today}

\begin{abstract}
We study the light-matter coupling of microcavity photons and an
interband transition in a one-dimensional (1D) nanowire. Due to the
Van Hove singularity in the density of states, resulting in a resonant
character of the absorption line, the achievement of strong coupling
becomes possible even without the formation of a bound state of an
electron and a hole. The calculated absorption in the system and
corresponding energy spectrum reveal anti-crossing behavior
characteristic of the formation of polariton modes. In contrast to the
case of conventional exciton polaritons, the formation of 1D Van Hove
polaritons will not be restricted to low temperatures and can be
realized in any system with a singularity in the density of states.
\end{abstract}

\maketitle

\section{Introduction}
Light-matter coupling is an area of research emerging at the
boundary between condensed matter physics and optics which has both
fundamental and applied dimensions. The possibility of reaching the
regime of strong coupling, for which confined cavity photons and
matter excitations are strongly mixed, is of particular
interest.\cite{KavokinBook} In this situation a new type of
elementary excitations, known as polaritons, appear in the system.
Having a hybrid nature, they combine the properties of both light
and matter. Several geometries have been proposed for the realization of
the strong coupling regime.

The interaction of a cavity photon mode with a two-level system
which mimics optical transitions in an individual atom or single
quantum dot (QD) is the origin of cavity quantum electrodynamics
(cavity QED).\cite{Dutra,Review_CQED} One should note that the
achievement of strong coupling in such a system is a non-trivial
task due to a rather small light-matter interaction constant.
However, recent advances in nanotechnology have led to the
possibility of creating high-finesse optical cavities and have
resulted in the observation of Rabi doublets and Mollow triplets in
the emission spectrum of individual
QDs.\cite{Laussy,Forchel,Hofling} Moreover, the strong coupling of a
single photon to a superconducting qubit, which has a two-level
structure as well, has been demonstrated in a radio-wave superconducting
cavity.\cite{Wallraff}

To increase light-matter coupling, quantum wells (QWs) can be used
instead of individual QDs. In this case, coupling occurs between a two-dimensional
(2D) QW exciton, associated with a sharp absorption peak slightly
below the bandgap energy, and a photonic mode of a planar cavity
tuned in resonance with it. Observed for the first time two decades
ago,\cite{Weisbuch} exciton-polariton physics is now experiencing increased
interest connected to the possible realization of polariton lasing with
an extremely low threshold,\cite{Imamoglu} and the achievement of
Bose-Einstein condensates\cite{Kasprzak} (BECs), and superfluid states
\cite{Amo} for temperatures much higher than for atomic
systems\cite{Anderson,Davis} and cold excitons in
solids.\cite{ButovNature} This is a consequence of the small
effective mass of polaritons which allows a pronounced manifestation
of quantum collective phenomena for a critical temperature around 20
K in CdTe structures \cite{Kasprzak} and even at room temperatures
in wide-band-gap materials with large exciton binding energy and
strong light-matter interaction (GaN,
ZnO).\cite{Christopoulus,Baumberg} Additionally, polaritons have been
proposed as basic ingredients for spinoptronic
devices\cite{PolaritonDevices} and all-optical logical elements and
integrated circuits.\cite{Liew,Circuit}

While most attention in the field of exciton polaritons is drawn to
two-dimensional structures, the strong light-matter interaction of
excitons in one-dimensional nanowires with a confined cavity mode has
also been studied.\cite{Kaliteevski,Chen} Their properties were shown to
be improved over the exciton-polaritons analogs in the quantum wells due 
to their larger exciton binding energy.\cite{Chen} However, the
physics of light matter coupling remains essentially the same.

Another system where strong light-matter coupling was experimentally
achieved is the intersubband transitions in quantum wells. It was
shown by Dini and co-workers\cite{Dini} that the absorption of
intersubband resonance placed into a cavity reveals the
characteristic anti-crossing behavior. The attractive peculiarity of
such a system in comparison to those based on conventional interband
exciton-polaritons is a non-vanishing ratio of the Rabi frequency to
the transition energy, which enables an exploration of the 
ultra-strong-coupling regime.\cite{Ultrastrong} In addition, unlike
interband transitions in a 2D system, strong electron-hole
interactions and the formation of excitons are not necessary for
obtaining the strong coupling regime,\cite{DeLiberato} although they
play a certain role in structures with highly doped QWs, where the
formation of intersubband plasmon-polaritons\cite{Geiser,Kyriienko}
and Fermi edge polaritons \cite{FermiEdge} can be observed.

For any experimental geometry, the main condition which must be
satisfied to drive the system into strong coupling is the presence
of a narrow resonance in its photoabsorption spectrum.
Mathematically, this condition reads
\begin{equation}
    g=\frac{\hbar\Omega_{R}}{2}\gg \gamma_{cav}-\gamma_{ex},
\end{equation}
where $g$ denotes the light-matter coupling strength, $\Omega_{R}$
is the corresponding  Rabi frequency (the light-matter interaction
constant), $\gamma_{cav}$ is the damping constant of the cavity mode
and $\gamma_{ex}$ is the width of the absorption
resonance.\cite{KavokinBook,Haug} In 2D interband absorption this
condition requires strong exciton-hole attraction resulting in the
formation of an exciton. However, in the 1D case this is not
strictly speaking necessary, since the behavior of the density of
states $\rho(E)$ in 2D and 1D is qualitatively different. In the
former case, the density of states is constant, while in the latter
it diverges around the points $E_0$ at which the energy as a
function of the momentum $k$ reaches its minimum,
\begin{equation}
\rho(E)\sim\frac{1}{\sqrt{E-E_0}}.
\end{equation}
This peculiarity, known as a Van Hove singularity, makes the optical
properties of 1D nanostructures different from those of the bulk and 2D cases and
leads to the resonant character of photoabsorption even without any
excitonic effects.

In this article we analyze the possibility of the realization of the
strong-coupling regime between a cavity photon and the interband
transition of a one-dimensional nanowire driven by the presence
of the Van Hove singularity in the 1D density of states. We consider
the case of the cavity mode tuned to the interband transition, while the 1D
exciton is detuned far from the resonance and thus does not affect
the properties of the system in the frequency range we consider. The
demonstrated spectrum of the elementary excitation reveals
anti-crossing behavior characteristic of the formation of
polaritonic modes for realistic cavity quality factors. We show that
this effect is robust against both finite temperature and
interaction corrections.

\section{Model}

\begin{figure}
\centering
\includegraphics[width=0.7\linewidth]{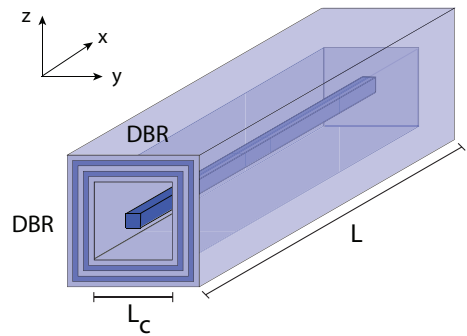}
\caption{(Color online) Sketch of the system. A semiconductor microcavity is formed with two pairs of distributed Bragg reflectors (DBRs)
with the cavity width in the $z$ and $y$ direction being $L_{c}$. The length of the 1D semiconductor wire is $L$.}
\label{Fig1}
\end{figure}

We consider a system consisting of a 1D semiconductor wire embedded
in an optical microcavity, which is tuned into resonance with a
direct interband transition (see the sketch in Fig. \ref{Fig1}).

To calculate the optical response of the coupled wire-resonator
system, we use an approach based on Green's function technique.
Information about the dispersion of elementary excitations in the
coupled quantum-wire--cavity system can be extracted from the poles
of the Green's function of the cavity photon interacting with
the interband transition $D(q,\omega)$. The absorption spectrum can be
determined from the polarization operator $\Pi(q,\omega)$. The
formalism we are going to use is analogous to that developed in
Ref.~[\onlinecite{Kyriienko}] for intersubband transitions.

\begin{figure}
\centering
\includegraphics[width=\linewidth]{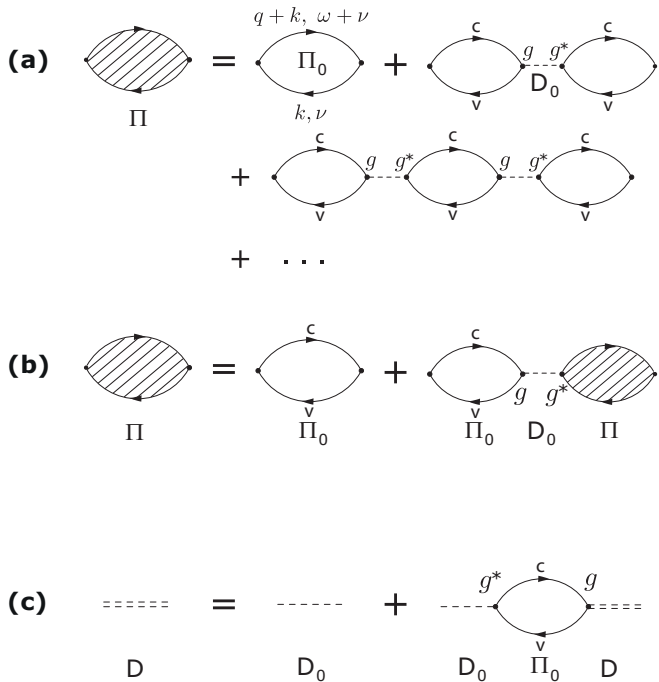}
\caption{(a) Feynman diagrams corresponding to the renormalized polarization operator $\Pi$, written as a sum of diagrams accounting for processes of multiple re-emissions and re-absorptions. $c$ and $v$ stand for the conduction and valence bands, respectively. (b) Dyson equation for the operator $\Pi(q,\omega)$. The vertex $g$ of the bubble diagrams denotes the coupling constant between the cavity photon and the interband excitation and the dashed line corresponds to the bare cavity photon Green's function $D_{0}$. (c) Dyson equation for the photon Green's function $D$, represented by a double dashed line, accounting for light-matter coupling.}
\label{Fig2}
\end{figure}

We start with a calculation of the polarization operator, accounting
for the possibility of multiple re-emissions and re-absorptions of a
cavity photon by the interband transition in the quantum wire. These
processes can be represented graphically as an infinite set of
diagrams  shown in Fig.~\ref{Fig2}(a), which can be reduced to the
Dyson equation shown in Fig. \ref{Fig2}(b). These series
diagrammatically describe the strong light-matter coupling regime,
where multiple interaction with the cavity mode renormalizes the overall
absorption in the system. The corresponding solution of the Dyson
equation reads
\begin{align}
\Pi(q,\omega)&=\frac{\Pi_{0}(q,\omega)}{1-g^{2}D_{0}(q,\omega)\Pi_0(q,\omega)},
\label{P}
\end{align}
where $\Pi_{0}$ denotes the interband polarization operator depicting
excited electron-hole pair in the conduction and valence bands and
$g$ stands for the light-matter interaction constant. $D_0$ is a
propagator of the cavity photon, given by \cite{Kyriienko}
\begin{align}
D_{0}(q,\omega)&=\frac{2\hbar \omega_{cav}(q)}{\hbar\omega^2-\hbar\omega_{cav}^2(q)+2i\gamma_{cav}\hbar \omega_{cav}(q)},
\label{D0}
\end{align}
with $\gamma_{cav}=h/\tau_{ph}$ representing the damping of the
cavity mode appearing due to the imperfectness of the cavity, where
$\tau_{ph}$ is the cavity photon lifetime. $\hbar\omega_{cav}$
denotes the dispersion of the cavity photon and can be written as
\begin{equation}
\hbar\omega_{cav}(q)=\frac{\hbar c}{n} |\mathbf{q}| = \frac{\hbar c}{n}\sqrt{q_x^2+q_y^2+q_z^2},
\end{equation}
where $n$ is the refractive index of the cavity and $c$ denotes the
speed of light. The confinement in the $y$ and $z$ directions leads to a
quantization of the photon momentum components $q_y$ and $q_z$, while it
can freely propagate in the $x$ direction. Here we are interested in the
photonic mode with the lowest energy which corresponds to the single
antinode lying in the center of the cavity. The dispersion of the cavity
photon as a function of $q_x$ (in the following we denote it as $q$
for simplicity) reads:
\begin{align}
&\hbar\omega_{cav}(q)=\frac{\hbar c}{n}\sqrt{q^2+\left(\frac{\pi}{L_c}\right)^2+\left(\frac{\pi}{L_c}\right)^2} \approx \frac{c\hbar L_c}{2\sqrt 2 \pi n}q^2\\ \notag
&+\frac{\sqrt 2 \pi c\hbar}{nL_c} \equiv\frac{\hbar^2q^2}{2m_{ph}}+E_g+\delta,
\end{align}
where $m_{ph}$ is the effective mass of the photon and $\delta$ is the detuning between the cavity mode and the bandgap $E_g$.

The interband polarization operator $\Pi_0$ used in Eq. (\ref{P})
describes the absorption of the wire in the absence of the cavity. Strictly
speaking, its calculation requires a full account of the many-body
interactions in the system and thus represents a formidable problem.
In the current work, however, we are interested in the effects of the
Van Hove singularity only, and as a first approximation we consider
non-interacting particles. However, in the later discussion we
will account for Coulomb interactions using the random phase
approximation (RPA) and compare the results with those of the non-interacting case.
The bare interband polarization operator $\Pi_0(q,\omega)$ is
represented by a ``bubble'' diagram and can be calculated as
\begin{align}
\notag
i\Pi_0(q,\omega)=&-\int\frac{dk}{2\pi/L}\frac{d\nu}{2\pi/\hbar}[iG_e(k+q,\nu+\omega)iG_h(k,\nu)\\ &+ iG_h(k+q,\nu+\omega)iG_e(k,\nu)].
\label{Pi0}
\end{align}
Here $G_e(q,\omega)$ and $G_h(q,\omega)$ are the Green's functions of an electron in the conduction band and a hole in the valence band, respectively,
\begin{align}
G_e(q,\omega)&=\frac{1}{\hbar\omega - E_{g}- \hbar^2q^2/2m_e+i\delta},\\
G_h(q,\omega)&=\frac{1}{\hbar\omega + \hbar^2q^2/2m_h-i\delta},
\end{align}
where $m_e$ and $m_h$ are the effective electron and hole masses
(both taken to be positive), and $\delta$ is an infinitesimal parameter. For the case of zero temperature an
analytical integration of Eq. (\ref{Pi0}) can be performed, which
leads to an explicit expression for the bare polarization operator
$\Pi_0$:
\begin{align}
\notag
\Pi_0(q,\omega)=\frac{L\sqrt{2\mu}}{\hbar}\left(\frac{f_1(q,\omega)}{\sqrt{\hbar\omega+E_g+\hbar^2q^2/M-i\gamma_{ex}}}-\right.\\ \notag
\\\left.-\frac{if_2(q,\omega)}{\sqrt{\hbar\omega-Eg-\hbar^2q^2/M+i\gamma_{ex}}}\right),
\end{align}
where $\mu$ denotes the reduced mass, $\mu^{-1}=m_e^{-1}+m_h^{-1}$,
$M=m_e+m_h$ and $L$ is the length of the wire. The functions
$f_1(q,\omega)$ and $f_2(q,\omega)$ are given by
\begin{align}
\notag
f_1(q,\omega)=&\frac{1}{\pi}\left[\text{tan}^{-1}\left(\frac{\pi/a +\beta_{e}q}{\sqrt{2\mu(\hbar\omega+E_g+\hbar^2q^2/M-i\gamma_{ex})}}\right) \right.\\
&+\left.\text{tan}^{-1}\left(\frac{\pi/a -\beta_{e}q}{\sqrt{2\mu(\hbar\omega+E_g+\hbar^2q^2/M-i\gamma_{ex})}}\right)\right],\\
\notag
f_2(q,\omega)=&\frac{i}{\pi}\left[\text{tanh}^{-1}\left(\frac{\pi/a +\beta_{h}q}{\sqrt{2\mu(\hbar\omega-E_g-\hbar^2q^2/M+i\gamma_{ex})}}\right) \right.\\
&+\left.\text{tanh}^{-1}\left(\frac{\pi/a -\beta_{h}q}{\sqrt{2\mu(\hbar\omega-E_g-\hbar^2q^2/M+i\gamma_{ex})}}\right)\right],
\end{align}
where we introduced the notations $\beta_{e}=m_{e}/M$ and
$\beta_{h}=m_{h}/M$ and $\gamma_{ex}$ denotes the non- radiative
lifetime of the excitation. The parameter $a$ defines the cut-off of the
integration at $\pm\pi/a$ and is proportional to the size of the
elementary cell of the material of the wire. For small momentum $q$,
the functions are close to unity in all frequency ranges.

The light-matter coupling constant $g$ can be estimated
as\cite{Koch}
\begin{equation}
g=|d_{cv}|\sqrt{\frac{\hbar\omega_{cav}}{2\epsilon\epsilon_{0}V}}\approx\sqrt{\frac{\hbar^2 e^2}{\epsilon\epsilon_{0}\mu L_c^2 L}},
\label{g}
\end{equation}
where $\epsilon$ and $V=L_c^2L$ are the dielectric permittivity and the 
volume of the cavity, respectively, and the cavity length parameters
$L_{c}$ and $L$ are shown in the geometry sketch (Fig.~\ref{Fig1}).
Note that $\Pi_0\sim L$ and $g\sim L^{-1/2}$, and the observable
quantities do not depend on the length of the system $L$.

The polarization operator $\Pi_0$ has both real and imaginary parts,
and the latter is related to the absorption coefficient of the
electron-hole interband excitation:\cite{Koch}
\begin{align}
\alpha(\omega)&=\frac{4\pi\omega}{n c}\chi''(\omega)\sim
    \text{Im}\Pi_0(q,\omega), \label{abs}
\end{align}
where $\chi''(\omega)$ is the imaginary part of the optical susceptibility.

The dispersion of the elementary excitations of the coupled 
quantum-wire--cavity system can be found from the poles of the renormalized
Green function of the cavity photon, accounting for the light-matter
coupling $D(q,\omega)$. The corresponding Dyson equation is shown in
diagrammatic form in Fig. \ref{Fig2}(c). Its solution gives
\begin{equation}
D(q,\omega)=\frac{D_0(q,\omega)}{1-g^{2}D_{0}(q,\omega)\Pi_{0}(q,\omega)},
\end{equation}
and the equation for the new eigenenergies of the system reads
\begin{equation}
1-g^{2}D_{0}(q,\omega)\Pi_{0}(q,\omega)=0. \label{Disp}
\end{equation}
\begin{figure}
\includegraphics[width=\linewidth]{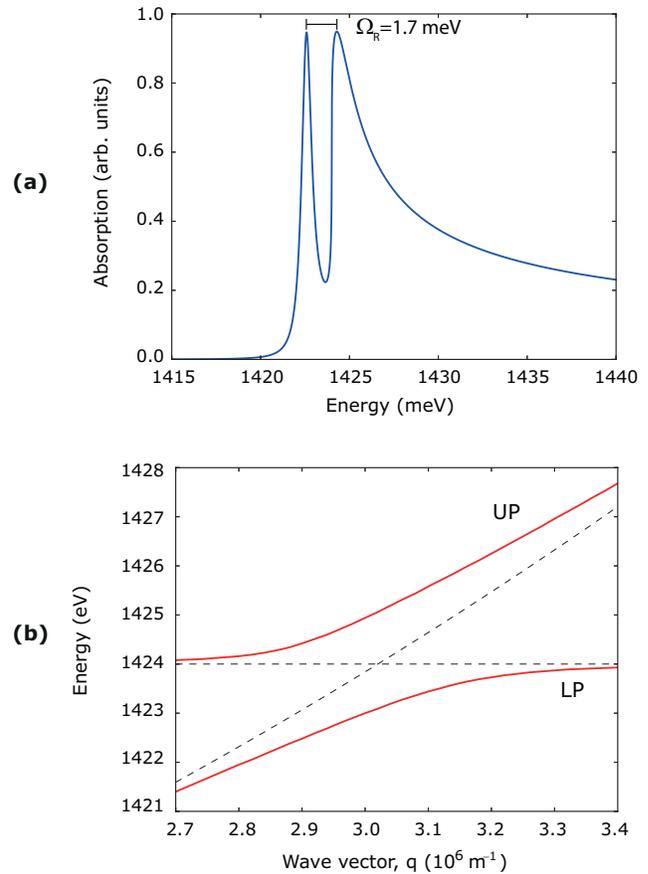}
\caption{(Color online) (a) An absorption plot showing polariton states formed by a microcavity photon and interband excitation in a 1D nanowire plotted for the $q$ corresponding to the anticrossing point. (b) The dispersion of elementary excitations in the system. Both plots are for one quantum wire and a photon lifetime of $\tau_{ph}=10$ ps, which corresponds to $\gamma_{cav}=0.4$ meV. These parameters yield a Rabi energy of $\hbar\Omega_R=1.7$ meV.}
\label{Fig3}
\end{figure}

\section{Results}
We calculate the renormalized polarization operator of the
cavity-excitation system using Eq. (\ref{P}), and find the energy
spectrum of the new modes. The absorption spectrum is calculated from
the imaginary part of $\Pi$ using Eq. (\ref{abs}), while the
dispersion relations are extracted from Eq. (\ref{Disp}). We used
standard parameters for a GaAs sample in our calculations.

Figure \ref{Fig3} shows the absorption spectrum of the coupled GaAs
quantum-wire--cavity system and the dispersion relation of the emergent
polaritonic modes. The lifetime of the cavity photon was taken as
$\tau_{ph}=10$ ps and the detuning of the cavity mode is
$\delta=-10$ meV. One can clearly see the anticrossing of the
eigenmodes, characteristic of the strong-coupling regime. The value
of the Rabi splitting for parameters considered here is about
$\hbar\Omega_R=1.7$ meV for a single quantum wire embedded in a
microcavity. The formation of polaritons is also revealed by a
double peak structure of the absorption spectrum shown in Fig.
\ref{Fig3}(a).
\begin{figure}
\includegraphics[width=0.85\linewidth]{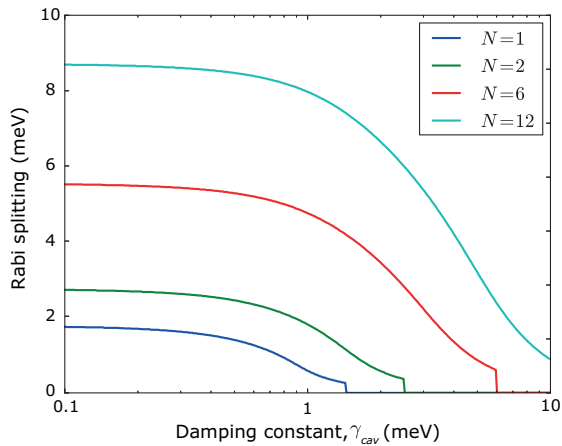}
\caption{(Color online) The Rabi splitting as a function of the cavity photon damping constant $\gamma_{cav}$ plotted for different numbers of quantum wires $N$ embedded in a semiconductor microcavity.}
\label{Fig4}
\end{figure}

The Rabi splitting can be enhanced by placing more than one wire in
the cavity. In this case the polarization operator corresponding to
a single wire, $\Pi_0$, should be replaced by $N_{QW}\Pi_0$, where
$N_{QW}$ is the number of wires in the cavity. The decrease of the
quality of the cavity, corresponding to the increase of the cavity
mode broadening $\gamma_{cav}$, results in quenching of the Rabi
splitting. The corresponding dependence for different values of
$N_{QW}$ is shown in Fig. \ref{Fig4}. The damping constant
$\gamma_{cav}$ scales from 0.1 to 10 meV on a logarithmic scale
which corresponds to lifetimes spanning from 40 to 0.4 ps, which
are possible to realize experimentally.

One can see that for $\gamma_{cav}\gtrsim 1$ meV, which corresponds
to a typical quality factor of semiconductor cavities, more than one
wire is required for reaching the strong-coupling regime.
However, with recent improvement in nanotechnology, fabrication of
high-quality resonators with $\gamma_{cav}\sim$ $0.1$ meV becomes
possible.\cite{Wertz} This in principle can allow the achievement of
strong coupling for an individual wire.

\section{Geometric, Finite-Temperature and Coulomb Corrections}

In the previous section we considered an idealized situation
corresponding to the case of zero temperature, neglecting
many- body effects and the placement of all the wires in the antinode of
the electric field of the cavity mode. In this section we consider
how deviations from these conditions affect the result.

\textit{Position of the wires.} Strictly speaking, the
$\sqrt{N_{QW}}$ enhancement of the Rabi splitting is valid only for
very thin wires placed exactly at the antinode of the cavity. For
 realistic systems we can calculate the Rabi splitting for an
array of quantum wires, accounting for the cavity mode structure.
For a single quantum wire placed in the center of a microcavity the
Rabi frequency is equal to $g=\hbar\Omega_R^{max}$. The index $max$
means that the value of the electric field is maximal in the center
of the cavity (the antinode for a $\lambda/2$ cavity). However, for a
square cavity it changes with deviation from the antinode position
as
\begin{equation}
\Omega_{R}(x)=\Omega_{R}^{max}\cos(\pi x/L_{c})\cos(\pi y/L_{c}).
\end{equation}
We can estimate the generalized Rabi splitting for an array of
nanowires as
\begin{equation}
\Omega_{R}^{\Sigma}=\sqrt{\sum_{i=1}^{N}\Omega_{R,i}},
\end{equation}
where $\Omega_{R,i}$ denotes the Rabi energy for each quantum wire.

\begin{figure}
\includegraphics[width=\linewidth]{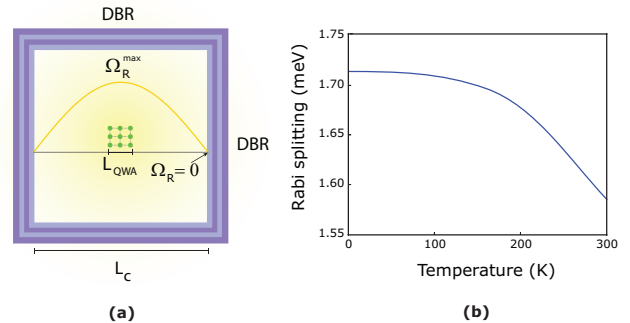}
\caption{(Color online) (a) Sketch of the quantum wire array (QWA) embedded in a cavity. The yellow line shows the behavior of the Rabi energy as a function of the confinement direction $x$. (b) Dependence of the Rabi splitting on temperature calculated for $\gamma_{cav}$ = 0.4 meV.}
\label{Fig5}
\end{figure}

We calculate $\Omega_{R}^{\Sigma}$ for $N_{QW}=9$ quantum wires for
the geometry sketched in Fig. \ref{Fig5}(a). The diameter of the
quantum wire is chosen as $L_{x}^{QW}=10$ nm with 15 nm separation,
and the GaAs cavity width is $L_c=400$ nm. The result gives
$\Omega_{R}^{\Sigma}=2.97\Omega_{R}^{max}$, which deviates by
several percent only from the simple estimation
$\Omega_{R}^{\Sigma}=\sqrt{N_{QW}\Omega_{R}^{max}}$.

\textit{Finite temperature.} To account for the finite temperature
of the system one can use the Matsubara representation of Green's
functions written in imaginary time.\cite{Bruus} The bare
interband polarization operator $\Pi_0$, accounting for
temperature effects, can be written as \cite{Koch}
\begin{align}
&\Pi_0(0,\omega)=2\int \frac{dk}{(2 \pi /L)}(f_{v,k}-f_{c,k})\Big[\frac{1}{\hbar\omega- E_g- \frac{\hbar^2k^2}{2\mu}+i\gamma_{ex}}\\
&-\frac{1}{\hbar\omega +E_g+ \frac{\hbar^2k^2}{2\mu}-i\gamma_{ex}}
\Big],
\end{align}
where $f_{v,k}$ and $f_{c,k}$ are the Fermi distribution functions
for the valence and conduction band, respectively. Here for
simplicity we account only for vertical transitions ($q \rightarrow
0$) and perform numerical integration on the momentum $k$. The Rabi
frequency can then be calculated as ib the zero temperature case
and is plotted as a function of the temperature in Fig.
\ref{Fig5}(b). One can see that with increasing temperatures up to
room temperature the Rabi frequency  decreases only slightly.
The system remains in the strong light-matter coupling regime
and 1D Van Hove polaritons are still observable. A similar
situation holds for intersubband polaritons.\cite{Dini}

\textit{Coulomb corrections.} Previously we restricted our
considerations to the non-interacting case. However, in real systems
Coulomb interactions do play a role and can change the spectrum of
elementary excitations in the system. Accounting for all possible
interactions is an extremely complicated task and usually can be
done only within certain approximations. For instance, accounting for
excitonic resonances requires the use of the ladder approximation.\cite{DasSarma99}
It leads to an integral Bethe-Salpeter equation and represents a formidable problem in itself.
We do not address this problem in the present work, assuming that
the exciton transition lies far from the bare interband transition and
thus does not affect the optical properties of the system in the
frequency range we consider.

\begin{figure}
\includegraphics[width=0.9\linewidth]{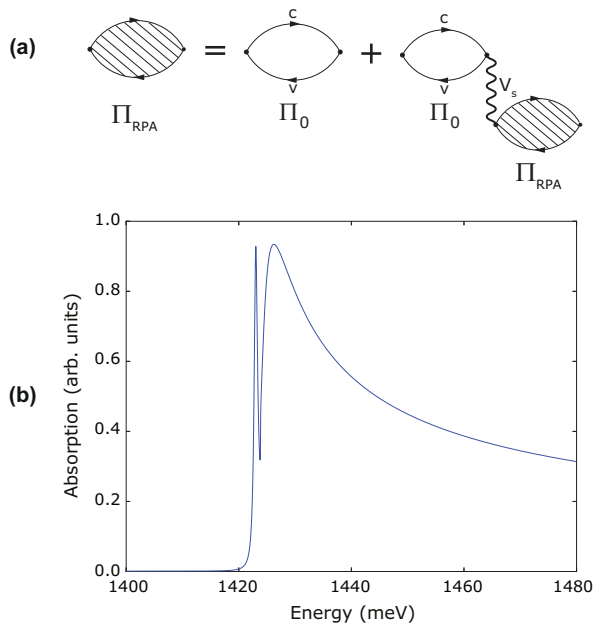}
\caption{(a) Feynman diagrams corresponding to RPA corrections to the polarization operator. The thick wiggly line $V_S$ corresponds to the screened Coulomb interaction. (b) Absorption spectrum of the cavity-wire system with Coulomb corrections taken into account using the RPA.}
\label{Fig6}
\end{figure}

To estimate the influence of the Coulomb corrections
we use the RPA approximation
\cite{DasSarma99,DasSarma96} shown in diagramatic form in Fig.
\ref{Fig6}(a). The resulting equation for the renormalized interband
polarization operator is given by
\begin{equation}
\Pi_{RPA}=\frac{\Pi_0}{1-V_{S}\Pi_0},
\end{equation}
where $V_{S}$ denotes the screened Coulomb interaction calculated in
Ref. [\onlinecite{DasSarma96}]. Following the same procedure as
before, we substitute for the interband polarization operator $\Pi_0$ a
modified one, $\Pi_{RPA}$, and find that Coulomb corrections in the
random phase approximation lead to a broadening and a slight shift of
the Van Hove singularity peak. However, the system remains in the
strong coupling regime, as can clearly be seen in Fig.
\ref{Fig6}(b).

Finally, we should note that semiconductor quantum wires can be
replaced by carbon nanotubes, which are commonly synthesized in
bundles\cite{Smalley96} or can be arranged in well-aligned
arrays.\cite{Wang10} In general, excitonic effects are very
important in semiconducting carbon nanotubes, as the presence of
strongly bound dark excitons results in luminescence
suppression.\cite{Kono07} However, tuning the cavity mode to the
nanotube's Van Hove singularity should significantly improve
light-matter coupling. In quasi-metallic nanotubes with small
curvature-induced band gaps and in metallic (armchair) nanotubes
with magnetic-field-induced gaps, excitonic effects can be
neglected;\cite{Hartmann11} so that our theory of Van Hove
polaritons becomes directly applicable. Narrow-gap carbon nanotubes
have recently attracted significant attention as promising
candidates for terahertz
applications.\cite{Kibis07,Portnoi08,Portnoi10} Metallic nanotubes
with magnetic-field-induced gaps are of a particular interest, since
their spectra can be easily tuned by an external magnetic field.
Their similarity to a two-level system is further enhanced by the
fast decrease of the dipole transition matrix element away from the
field-induced band gap.\cite{PortnoiShelykh09} Another system, for
which the developed theory is highly relevant is a bulk
semiconductor with Van Hove singularities resulting from
quasi-one-dimensional motion along a quantizing magnetic field.
Carbon nanotubes in microcavities in both the optical and terahertz
frequency ranges as well as bulk materials with
magnetic-field-induced Van Hove singularities are subjects of our
future work.

\section{Conclusions}
In conclusion, we have studied the light-matter coupling of a
microcavity photon and an interband transition in a 1D nanowire. Due to
the resonant character of the absorption spectrum provided by the
Van Hove singularity of the 1D density of states, the achievement of
the strong coupling regime becomes possible even in the absence of
excitonic effects. We have calculated the dispersions of the resulting
polariton modes and the absorption spectra of the coupled
wire-cavity system for realistic values of parameters. We have examined
the influence of Coulomb corrections and have shown that 1D Van Hove
polaritons are robust against temperature changes and can exist even
at room temperature.

\section*{Acknowledgments}
This work was supported by the EU FP7 ITN Spin-Optronics (Grant No. 237252) 
and the FP7 IRSES projects SPINMET (Grant No. 246784) and QOCaN (Grant No.
316432). O. K. acknowledges support from the Eimskip Fund.

\end{document}